\documentclass[aps,twocolumn, superscriptaddress, floatfix, longbibliography, svgnames, pra]{revtex4-2}
\usepackage[T1]{fontenc} 
\usepackage{tikz,tikzscale, relsize, pgfplots, xcolor, amssymb, amsmath, pgfplots, tikz-3dplot, float}
\usepackage{slashbox}
\usetikzlibrary{backgrounds, arrows, calc, decorations, positioning, decorations.pathmorphing,decorations.pathreplacing, 
	decorations.markings, fixedpointarithmetic, fit, patterns, decorations.text, backgrounds,matrix,plotmarks, spy}
\usepgfplotslibrary{fillbetween}
\pgfplotsset{compat=1.17}
\usetikzlibrary{external}
\newcommand{\Aa}{\mbox{\normalfont\AA}}
\renewcommand{\fnum@table}{TAB. \thetable}
\begin{document}
	\title{Optical and acoustic plasmons in the layered material Sr$_2$RuO$_4$}
	
	\author{J. Schultz}
	\email{j.schultz@ifw-dresden.de}
	\affiliation{Leibniz Institute for Solid State and Materials Research Dresden, Helmholtzstra{\ss}e 20, 01069 Dresden, Germany}
	
	\author{A. Lubk}
	\affiliation{Leibniz Institute for Solid State and Materials Research Dresden, Helmholtzstra{\ss}e 20, 01069 Dresden, Germany}
	\affiliation{TU Dresden, Institute of Solid State and Materials Physics, Haeckelstra{\ss}e 3, 01069 Dresden, Germany}
	
	\author{F. Jerzembeck}
	\affiliation{Max Planck Institute for Chemical Physics of Solids, Nöthnitzer Stra{\ss}e 40, D-01187 Dresden, Germany}
	
	\author{N. Kikugawa}
	\affiliation{National Institute for Materials Science, Tsukuba 305-0003, Japan}
	
	\author{M. Knupfer}
	\affiliation{Leibniz Institute for Solid State and Materials Research Dresden, Helmholtzstra{\ss}e 20, 01069 Dresden, Germany}
	
	\author{D. Wolf}
	\affiliation{Leibniz Institute for Solid State and Materials Research Dresden, Helmholtzstra{\ss}e 20, 01069 Dresden, Germany}
	
	\author{B. B\"uchner}
	\affiliation{Leibniz Institute for Solid State and Materials Research Dresden, Helmholtzstra{\ss}e 20, 01069 Dresden, Germany}
	\affiliation{TU Dresden, Institute of Solid State and Materials Physics, Haeckelstra{\ss}e 3, 01069 Dresden, Germany}
	
	\author{J. Fink}
	\email{j.fink@ifw-dresden.de}
	\affiliation{Leibniz Institute for Solid State and Materials Research Dresden, Helmholtzstra{\ss}e 20, 01069 Dresden, Germany}
	\affiliation{TU Dresden, Institute of Solid State and Materials Physics, Haeckelstra{\ss}e 3, 01069 Dresden, Germany}

	\begin{abstract}
		
	\end{abstract}

	\date{\today}
	
	\maketitle
	
	\section{Abstract}
	{\bf We use momentum-dependent electron energy-loss spectroscopy in transmission to study collective charge excitations in the 
		layer metal Sr$_2$RuO$_4$. This metal has a transition from a perfect Fermi liquid below \textit{T}\,$\approx$\,30 K into a "strange" metal phase above \textit{T}\,$\approx$\,800 K.
		We cover a complete range between in-phase and out-of-phase oscillations. Outside the classical range of electron-hole excitations, leading to a Landau damping, we observe well-defined plasmons. The optical (acoustic) plasmon due to an in-phase (out-of-phase) charge oscillation of neighbouring layers exhibits a quadratic (linear) positive dispersion. Using a model for the Coulomb interaction of the charges in a layered system, it is possible to describe the  range of optical plasmon excitations at high energies in a mean-field random phase approximation without taking correlation effects into account. In contrast, resonant inelastic X-ray scattering data show at low energies an enhancement of the acoustic plasmon velocity due to correlation effects. This difference can be explained by an energy dependent effective mass which changes from $\approx$ 3.5 at low energy to 1 at high energy near the optical plasmon energy. There are no signs of over-damped plasmons predicted by holographic theories.
	}
	
	\section{Plasmons in "strange" metals}\label{section:intro}
	"Strange" metals are at present one of the most interesting research fields in solid state physics \cite{Zaanen2019}. Due to the strong interaction between their charge carriers, they show a deviation from the Fermi-liquid behaviour, e.g., they do not show a quadratic but a linear temperature  dependence of the electrical resistivity or linear in energy scattering rate in Angle-Resolved Photoemission-Spectroscopy (ARPES)]~\cite{Cooper2009,Valla1999}. Moreover, there is no saturation at the Mott-Ioffe-Regel limit~\cite{Hussey2004}. The unconventional and in some cases high-temperature superconductivity detected in these materials is supposed to be related to their "strange" normal-state electronic structure. Doped cuprates are prototypes of these "strange" metals. The non-Fermi liquid properties could be explained by a  continuum of excitations up to an ultraviolet cutoff frequency $\omega_\text{c}$ in the electronic susceptibility~\cite{Varma1989}, leading to a phenomenological marginal Fermi liquid theory. Integrating over this continuum yields a linear in energy imaginary part of the self-energy or scattering rate~\cite{Norman2003}. Electron Energy-Loss Spectroscopy (EELS) is a suitable experimental method to verify the existence of such a continuum because it measures the imaginary part of the electronic susceptibility $\Im [\chi({\bf q},\omega)]$.
	Here ${\bf q}$ is the momentum and $\omega$ is the energy.
	In nearly-free electron metals, there exists a $2k_{\text{F}}$ (with $k_{\text{F}}$ equal to the Fermi wave vector) wide stripe of a continuum of particle-hole excitations~\cite{Lindhard1954} which starts at $q \approx \omega /v_\text{F}$, where $v_{\text{F}}$ is the Fermi velocity.
		In simple metals collective excitations (plasmons) exist below the critical wave vector $q_\mathrm {crit}$ $\approx \omega_{\text{P}} /v_{\text{F}}$ which is determined by the plasmon energy $\omega_\text{P}$ and the Fermi velocity $v_{\text{F}}$. Above this momentum  the plasmons merge into the continuum and therefore are (Landau) damped by a decay into particle-hole excitations.  Usually, $q_\mathrm{crit}$ is between half and one \AA $^{-1}$.
	
	\par
	Surprisingly, early transmission EELS (T-EELS) studies of the highly correlated doped cuprates, using dedicated T-EELS spectrometers~\cite{Raether1980,Schnatterly1979,Fink1989} with high momentum resolution, showed this behavior with a weak damping of plasmons below $q_\mathrm{crit}$~\cite{Nuecker1989,Wang1990, Romberg1990,Nuecker1991,KNUPFER1994,Roth2020}. On the other hand, T-EELS studies using transmission electron microscopes with weak momentum resolution~\cite{Terauchi_1995,Terauchi1999} showed no plasmon but a  continuum. The difference can be easily explained by the poor momentum resolution in the TEM experiments ($ \Delta q =$ 10 and 30 \AA $ ^{-1}$) which averages over the whole Brillouin zone (BZ) and therefore measures predominantly the Landau continuum above $q_\mathrm{crit}$.
	
	\par
	At variance with the early T-EELS measurements of collective excitations in hole doped cuprates
	great attention attracted the prediction of over-damped plasmons in "strange" metals and a replacement of these excitations by a continuum~\cite{Romero-Bermudez2019}.
	The authors concluded that a new kind of theory of strongly interacting matter may be needed to explain this. The latter may also be connected to the phenomenon of high-$T_\text{c}$ superconductivity. The over-damped plasmon was discussed in terms of holographic theories, which predicted, different from the classical Landau damping, a strong damping even for the long wavelength plasmons due to quantum critical fluctuations. Recently, this work was supported by similar calculations~\cite{eede2023}. 
	
	\par
	Furthermore, there are several recent experimental EELS experiments in reflection (R-EELS), supporting the theories which predict over-damped plasmons in "strange" or highly correlated metals.
	Only at very small momenta a well-defined plasmon exists followed by a transition into a featureless momentum-independent constant-in-frequency continuum well below $q_\mathrm{crit}$~\cite{Mitrano2018,Husain2019,Husain2023,Chen2024}.
	Moreover, there is a very recent ARPES study on doped cuprates in which these holographic theories are supported by an asymmetric line shape at higher energies~\cite{Mauri2024}.
	
	\par
	There are other differences between T-and R-EELS results from hole-doped cuprates: At small $q$, i.e. long wave length, T-EELS data show a positive dispersion which can be explained in RPA using an unrenormalised band structure~\cite{Nuecker1989,Wang1990,Nuecker1991,Grigoryan1999}. On the other hand, R-EELS data show a negative dispersion, which may indicate  a more localised electron liquid. The hybridisation  of the d-bands with the s-band in the alkali metals or many-body effects, when moving from Na to Cs is supposed to turn the plasmon dispersion from positive to negative~\cite{Felde1989a,Fleszar1997}. 
	On the other hand, the different result between T-EELS and R-EELS possibly can be explained by different response functions with respect to surface and bulk properties.
	\par
	In this context, we mention that  in various cuprates, well pronounced dispersive acoustic plasmons were detected by resonant inelastic X-ray scattering (RIXS) ~\cite{Hepting2018,Nag2020,Singh2022}. In all these measurements, weakly damped plasmons were detected for momentum ranging up to half of the size of the BZ.
	
	\par
	For understanding the differences between T- and R-EELS on cuprates and to understand the influence of correlation effects on collective charge oscillation in general, we present here T-EELS data on the related metal Sr$_2$RuO$_4$~\cite{Maeno1994,Bergemann2003}. It is in some way intermediate between a normal Fermi liquid metal and a strange metal. Below $\approx$ 30 K it is a perfect Fermi liquid which transforms at low temperatures $T_\mathrm{c} =$ 1.5 K into an unconventional superconductor \cite{Mackenzie1998}. It has other similarities to the cuprates: it has a perovskite structure  formed by transition metal oxides layers. The essentially 2D correlated electronic structure is formed  by three bands and has a van Hove singularity close to the Fermi level.

	Deviating from the cuprate high $T_\mathrm{c}$ superconductors, it is a stoichiometric compound without crystallographic disorder due to dopant ions. Furthermore, the temperature dependence of the transport properties are more complicated. Above $ T\approx $ 30 K there is a crossover region in which Sr$_2$RuO$_4$ exceeds at $T_\mathrm{MIR} \approx $ 800 K the Mott-Ioffe-Regel limit, i.e., it turns to a "bad metal". These transport properties are partially related to Hund's rule coupling which causes strong correlation effects far from the insulating state~\cite{Medici2011}.
	
	\par
	Recently, we have studied the electronic structure of Sr$_2$RuO$_4$ by an investigation of the optical plasmon excitations with momenta parallel to the layers~\cite{Knupfer2022} using a dedicated T-EELS spectrometer~\cite{Fink1989}. Also in this highly correlated material, a well-defined plasmon could be detected near 1.5 eV. The plasmon has a {\textit{positive}} dispersion and decays into a continuum of particle-hole excitations due to Landau damping, which could be explained in the framework of the random phase approximation (RPA) using an {\textit{unrenormalised}} band structure.
	\par
	Most of the previous momentum dependent EELS studies on layered materials were performed for a wave vector parallel to the layers. The reason for this is that thin samples (T-EELS) or clean surfaces (R-EELS) are easily prepared by a cleavage of the crystals parallel to the layers.
	In the present work, by focused ion beam milling, we are able to prepare thin electron transparent lamella in which the layers are perpendicular to the surface. Using such samples, it was for the first time possible to map out a complete set of plasmon excitations with momentum between parallel and perpendicular to the layers almost in the entire BZ. In this way, it is possible to control theoretical work on plasmon excitations in layered compounds, which is available since many decades~\cite{Grecu1973,Fetter1974,DasSarma1982,Greco2016}. 
	\par
	At present, there is a strong discussion, whether spectroscopic results on the damping and dispersion of plasmons can be explained on the basis of mean field theories such as RPA or whether we need new theories to explain valence band EELS results (see also the recent EELS review~\cite{Abbamonte2024} which contrast the conflicting results of T-EELS and R-EELS). The present article strongly supports the results derived from T-EELS.
	
	\section{Plasmon dispersion in layered systems}\label{section:plasdis}
	The dynamic structure factor is determined by the Fourier transformation of the charge density-density correlation function. It can be expressed~\cite{Platzman1973} by
	the dynamical susceptibility $\chi({\bf q},\omega)$:
	\begin{equation}\label{eq:dis}
		S({\bf q},\omega)\propto \Im \left[\chi({\bf q},\omega)\right]\propto \Im\left[ -\frac{1}{\epsilon({\bf q},\omega)}\right].
	\end{equation}
	Here, $\epsilon({\bf q},\omega)$ is the complex dielectric function.
	\par
	The calculation of the Lindhard-Ehrenreich-Cohen susceptibility of the many-body system of the charge carriers in solids is a challenging task. 
	The susceptibility $\chi_0$ for a non-interacting one-band electron liquid is given by~\cite{Ehrenreich1959,Nuecker1991}:
	\begin{equation} \label{eq:Ehren}
		\chi_0 (\omega,{\bf q}) = \int_\mathrm{BZ}M({\bf q,k})\frac{ 2F({\bf k})\Delta E({\bf q,k})}{(\omega+i\Gamma)^2-\Delta E^2({\bf q,k})}d^3k\,.
	\end{equation}
	Here $\Delta E=E_{\bf k+q}-E_{\bf k} $, $E_{\bf k}$ are the band energies of the electrons having a momentum $\bf k$, $M({\bf q,k})$ is related to matrix elements, $\Gamma$ is the lifetime broadening of the particle-hole excitations, and $F({\bf k})$ is the Fermi function.
	\par
	While $\chi_0$ is the Lindhard-Ehrenreich-Cohen susceptibility for single-particle excitations related to an external field, stemming from the field of the scattering electron, $\chi$ is the susceptibility for the total field, including the induced one. 
	Running a self-consistency cycle, we obtain in the
	mean field RPA the  result:
	\begin{equation}\label{eq:RPA}
		\chi^\mathrm{RPA}({\bf q},\omega)=\frac{\chi_0({\bf q},\omega)}{\epsilon_\text{b}-V({\bf q})\chi_0({\bf q},\omega)}.
	\end{equation}
	
	Here, V({\bf q}) is the Fourier transformed Coulomb interaction between the charge carriers and $\epsilon_\text{b}$ is the background dielectric constant.
		The dielectric function can be calculated from
		\begin{equation}\label{eq:diel}
			\epsilon({\bf q},\omega)=\epsilon_b-V({\bf q})\chi_0({\bf q},\omega).
		\end{equation}
		$\epsilon_{\text{b}}$ is the background dielectric function.
	
	\par
	In this approximation and for small damping, there are in addition to the single-particle excitations collective excitations, termed plasmons. 
	The energy of the plasmon is determined by the zeros of the real part of the denominator of Eq.~(\ref{eq:RPA}). The long wavelength energy of the plasmon in the RPA is given by
	\begin{equation}\label{eq:plasmon}
		\omega_\mathrm{P}(0)^2=\frac{4\pi N e^2 }{\epsilon_\text{b} (m^*/m_{0})}
	\end{equation}
	with $N$ being the density of the charge carriers, $m^*$ the effective mass, $m_{0}$ the unrenormalised mass,  and $e$ the elementary charge.
	
	\par
	For small but finite momentum, up to $q^2$ and within the RPA, the  dispersion  is given by 
	\begin{equation}\label{eq:disp}
		\omega_\mathrm{P}(q)=\omega_\mathrm{P}(0)+A_\mathrm{RPA}q^2+ .....;\hspace{0.5 cm}
		A_\mathrm{RPA}= ( A_1 + A_2).
	\end{equation}
	
	$A_1$ is related by the finite compressibility or the squared averaged Fermi velocity of the electron liquid and is always positive.  
	$A_2$, which is always negative, is proportional to the size of the effective mass~\cite{Grigoryan1999}.

	For free-electron metals only 
	\begin{equation}\label{eq:A1}
		A_1=\frac{1}{\omega_{\text{P}}(0)} \frac{3}{10}\left<v_\mathrm{F}^2\right>_{\bf q}
	\end{equation}
	determines the optical plasmon dispersion.
	Here, the averaged squared Fermi velocity along the ${\bf q}$ direction is defined by~\cite{Nuecker1991}
	\begin{equation}\label{eq:vF2}
		\left<v_\mathrm{F}^2\right>_\mathrm{{\bf q}}=\left<\left(\frac{\bf q}{\hbar q}  \frac{\partial E_{\bf{k}}}{\partial \bf{k}}\right)^2\right>.
	\end{equation}
	For metals where the band dispersion is strongly reduced by a finite effective mass $m^*/m_0$, the negative $A_2$  may dominate the optical plasmon dispersion, leading in total to a negative dispersion~\cite{Grigoryan1999}.
	\par
	In the following we discuss the structure factor V({\bf q}).
	In a homogeneous electron system
	\begin{equation}\label{eq:hom}
		V({\bf q})=\frac{4 \pi e^2}{q^2}\,.
	\end{equation}
	For a system, built up by 2D layers separated by the distance $d$ we use the Fetter model with the Coulomb potential 
	\begin{equation}\label{eq:lay}
		V({\bf q})=V(q_{||},q_{\perp}) =\frac{4 \pi e^2}{q^2}\frac{q_{||} d }{2} \frac{\mathrm{sinh}(q_{||} d)}{\mathrm{cosh}(q_{||} d)-\mathrm{cos}(q_{\perp} d)},
	\end{equation}
	where $ q_{||}~(q_{\perp}) $ is the momentum parallel (perpendicular) to the layer~\cite{Fetter1974}.
	Therefore, in a layered system, the plasmon dispersion is not only determined by the compressibility and by the effective mass of the electron liquid (see Eq.~\ref{eq:disp}) but also by the structure factor $V({\bf q)}$, depending on $q_{||}d$ and $q_{\bot} d$
	\begin{equation}\label{eq:disfet}
		\omega_\mathrm{P}(q)^2=s^2q^2+\omega_\mathrm{P}(0)^2\frac{q_{||} d }{2} \frac{\mathrm{sinh}(q_{||} d)}{\mathrm{cosh}(q_{||} d)-\mathrm{cos}(q_{\perp} d)}
	\end{equation}
	with $s^2=\frac{1}{2}\left<v_\mathrm{F}^2\right>_\mathrm{{\bf q}}$.
	For a dielectric function that is consistent with the above dispersion relation we refer to the Method section. For small $q_{||}$ and $q_{\perp} d= 0$, the structure factor $V(q_{||},q_{\perp})$ is the same as in the homogeneous electron gas. In this case, the charge oscillations in the layers are in phase and the plasmon dispersion is the same as in a homogeneous 3D electron system. For $q_{\perp} d= \pi$ the charge oscillations between neighboring layers are out of phase. 
	\par
	This leads to an acoustic plasmon, with a rather small energy gap due to the layer interaction~\cite{Hepting2022} in the long wavelength limit.
	The dispersion of the acoustic plasmon for small $q_{||}$ is given by
	\begin{equation}
		\label{eq:acoust}
		\omega_\mathrm{p} (q_{||}) = q_{||} \sqrt{s^2+\omega_\mathrm{p}(0)^2 \frac{ d^2 }{4}}.
	\end{equation}
	Very often, the first term in the square root is considerably smaller than the second. Then the phase velocity of the acoustic plasmon along the direction of $q_{||}$ is given by
	\begin{equation}
		\label{eq:velocity}
		v_\mathrm{p} = \frac{\omega_\mathrm{p}(0) d}{2}.
	\end{equation}
	Besides the hydrodynamic Fetter model, the plasmon dispersion of a layered system was also derived by means of RPA, leading to similar results~\cite{Apostol1975}.
	\section{EELS data}
\begin{figure}[ht]
		\centering
		\includegraphics[width=.475\textwidth]{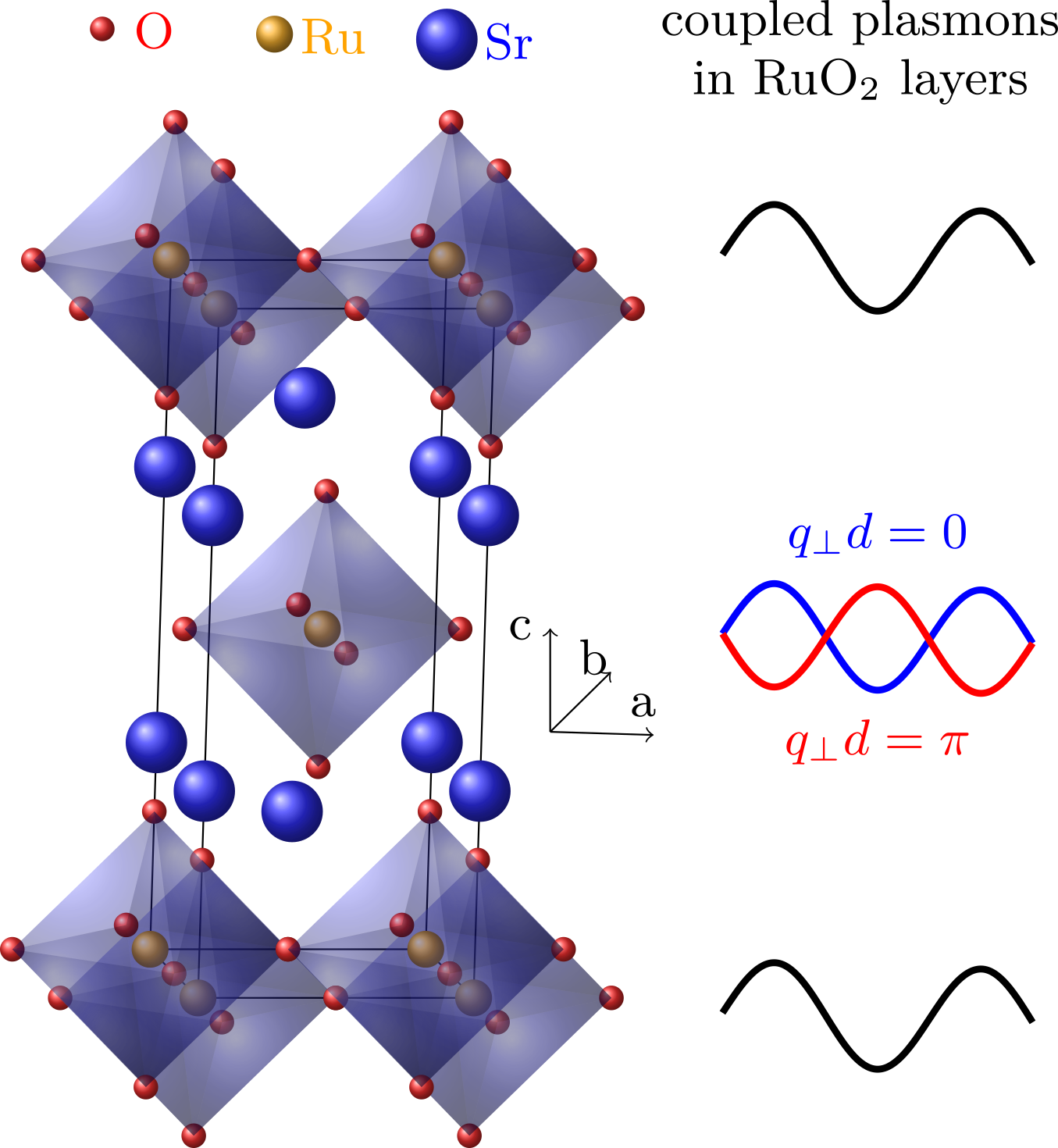}
		\caption{Crystal structure of Sr$_2$RuO$_4$ and schematic representation of in-phase and out-of-phase two dimensional charge oscillations}
		\label{fig:fig1}
	\end{figure}

\begin{figure}[ht]
		\centering
		\includegraphics[width=.475\textwidth]{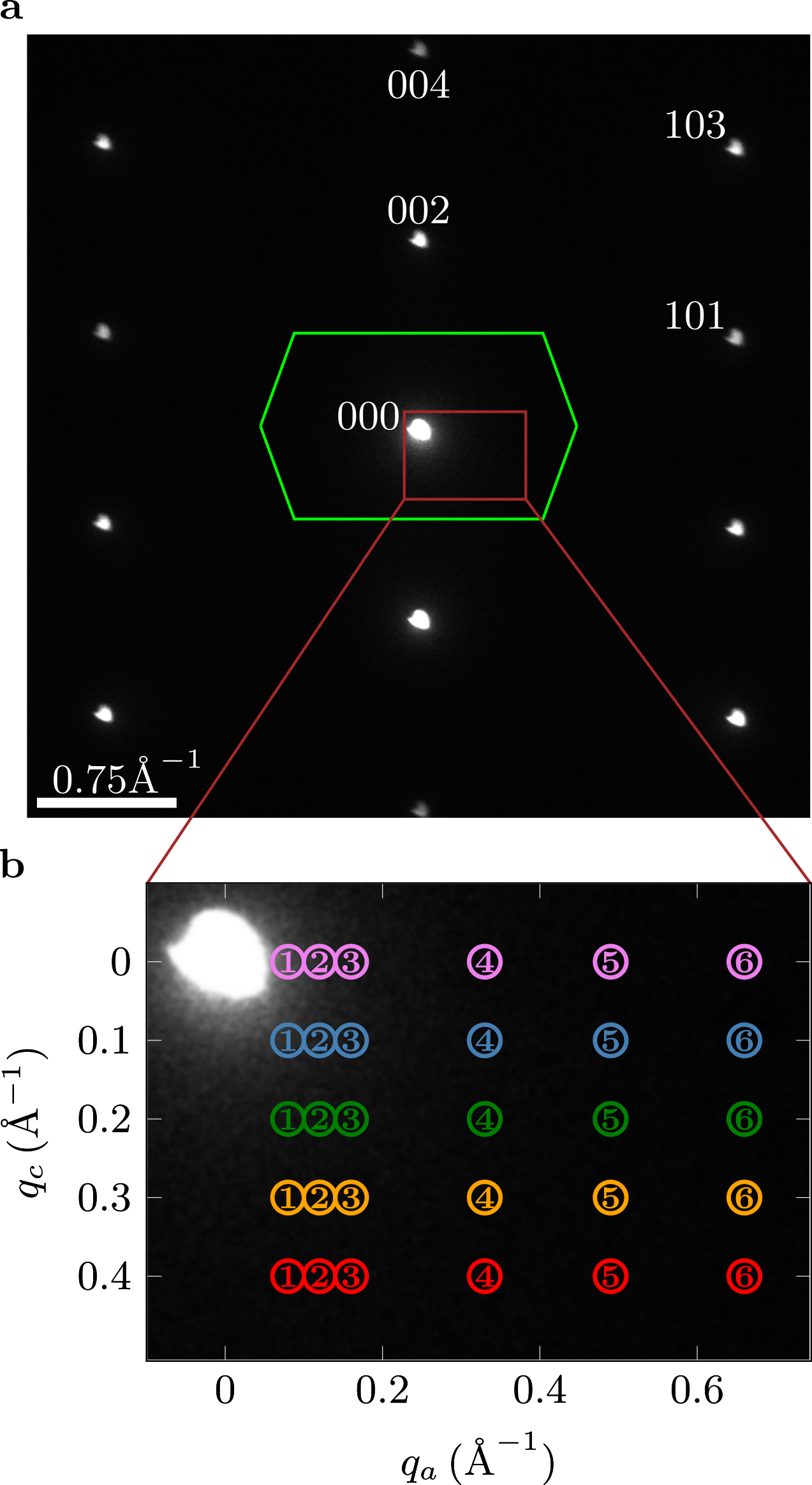}
		\caption {(a) Indexed electron diffraction pattern (white dots) in the ($q_{||},q_\perp)$ plane, including the reduced Brillouin zone (green) for equal layers along the ${\bf c}$-axis.
			(b) Brown momentum range [see (a)] in which loss spectra are recorded for various
			$q_{||}$ and $q_\perp$ values. The colours purple, blue, green, yellow, and red correspond to $q_\perp$
			equal 0, 0.1, 0.2 0.3, and 0.4 \AA $^{-1}$, respectively. The diameter of the filter entrance aperture [indicated by the colored circles in (b)] defining the momentum range in one EEL spectrum (momentum resolution) corresponds to a momentum of 0.04 \AA $^{-1}$. }
		\label{fig:fig2}
	\end{figure}

	T-EELS measurements were performed on single-crystalline Sr$_2$RuO$_4$  at room temperature. Fig.\,\ref{fig:fig1} shows the crystal structure of Sr$_2$RuO$_4$ which consists of RuO$_2$ layers stacked with SrO spacer layers along the ${\bf c}$-axis direction.
	The thin films were characterised by \textit{in-situ} electron diffraction and the crystallographic axes were oriented with respect to the $({\bf a,c})$ plane [see Fig.\,\ref{fig:fig2} (a)]. 
	The distance $d =$ 6.36~\AA\, between the RuO$_2$ layers is half the $ \bf c$-axis lattice constant. The thin lamellas for T-EELS with a normal direction parallel to the ${\bf b}$-axis were cut with a focused ion beam. In this way, EELS experiments were possible with the momentum parallel to the ${(\bf a,c})$-plane. We emphasize that this momentum range is different from our previous EELS study on Sr$_2$RuO$_4$~\cite{Knupfer2022} where we covered
	the momentum range in the (${\bf a}$,${\bf b}$)-plane.
	
	\par
	T-EELS was performed with a primary electron energy of 80 keV and with an energy and momentum resolution of 120 meV and 0.04 \AA$^{-1}$, respectively. The momentum-resolved EELS data were sequentially recorded in the ($q_{||}=q_{\text{a}},q_\perp$) momentum plane as depicted in Fig.\,\ref{fig:fig2}(b). In Fig.\,\ref{fig:fig3} we show typical EELS intensities as a function of the energy for various $q_{||}$ and $q_\perp$ values. With increasing $q_{||}$ the plasmon energy slightly increases, whereas for increasing $q_\perp$ the plasmon energies decrease.
	At low total momentum it is difficult to see a well defined plasmon due to the high intensity of the quasi-elastic peak. The same holds for high total momentum because the T-EELS cross section is decreasing with $1/q^2$ (see below). Except for the described cases, well developed and dispersing plasmon excitations below 1.8 eV are visible. From the fit to the loss data with a Drude function,  we obtain the energy, the width, and the intensities (see Methods).
	\par
	The energy is plotted in Fig.\,\ref{fig:fig4} as a function of $q_{||}$ for various $q_\perp$ values together with theoretical curves calculated in the framework of the Fetter model (see Section \ref{section:plasdis}). 
	We use the parameters $\omega_\mathrm{p}$(0) = 1.48 eV (from optical spectroscopy~\cite{Stricker2014}, $d=$ 6.36 \AA, and $\left<v_\text{F}^2\right>_{100}$ = 
	4.91 (eV \AA)$^2$.
	The latter value for the three Ru 4d $t_{2g}$ bands crossing the Fermi level was derived from a tight-binding (TB) band structure~\cite{Liebsch2000} (see Methods).
	For all plasmon energies above the continuum, within error bars, there is rather good agreement between theory and experiment. There is a continuous transition between the optical plasmon for $q_\perp= 0$ (purple data) at higher energy to the acoustic plasmon ($q_\perp=$ 0.4 \AA$^{-1}$) close to $q_\perp = \pi/d =0.49$ \AA $^{-1}$ (red data) at lower energy. Due to our finite energy resolution we cannot follow the acoustic plasmon to zero energy. Furthermore the momentum resolution of the instrument is limited by the finite width of the collection aperture of our spectrometer. Since the signal is integrated over the latter we observe a drop of the plasmon energy also for the optical plasmon near $q_\perp=0$ (see purple data in Fig.~\ref{fig:fig4}).
	\par
	The difference between extrapolated RPA plasmon dispersion and experimental data for $q_{||} \ge q_\mathrm{crit} $ (see Fig.~\ref{fig:fig4}) can be explained by Landau damping. For these wave vectors, the plasmon decays into  over-damped plasmon excitations and into  spectral weight which is caused by the Lindhard continuum. This leads to a reduction of the energy of the maximum in the total loss function. This reduction  has been observed in our calculations presented in Ref.~\cite{Knupfer2022}. A similar behavior has been also observed for Al~\cite{Batson1983}.
	In Fig.\,\ref{fig:fig4} we also show the continuum of the  single-particle intra-band transitions $\chi_0$ (see the grey region) calculated using an unrenormalised TB band structure~\cite{Liebsch2000} (see Methods).
	The optical plasmon merges into the continuum near $q_{||}\approx$ 0.4 \AA $^{-1}$. We assign this value to the critical momentum $q_{\mathrm{crit}}$. This value is further corroborated by the rapid increase of the plasmon width at this wave vector (see Fig.\,\ref{fig:width}).
	
	\section{Optical and acoustic plasmons in layered "strange" metals}
\begin{figure*}
		\centering
		\includegraphics[width=\textwidth]{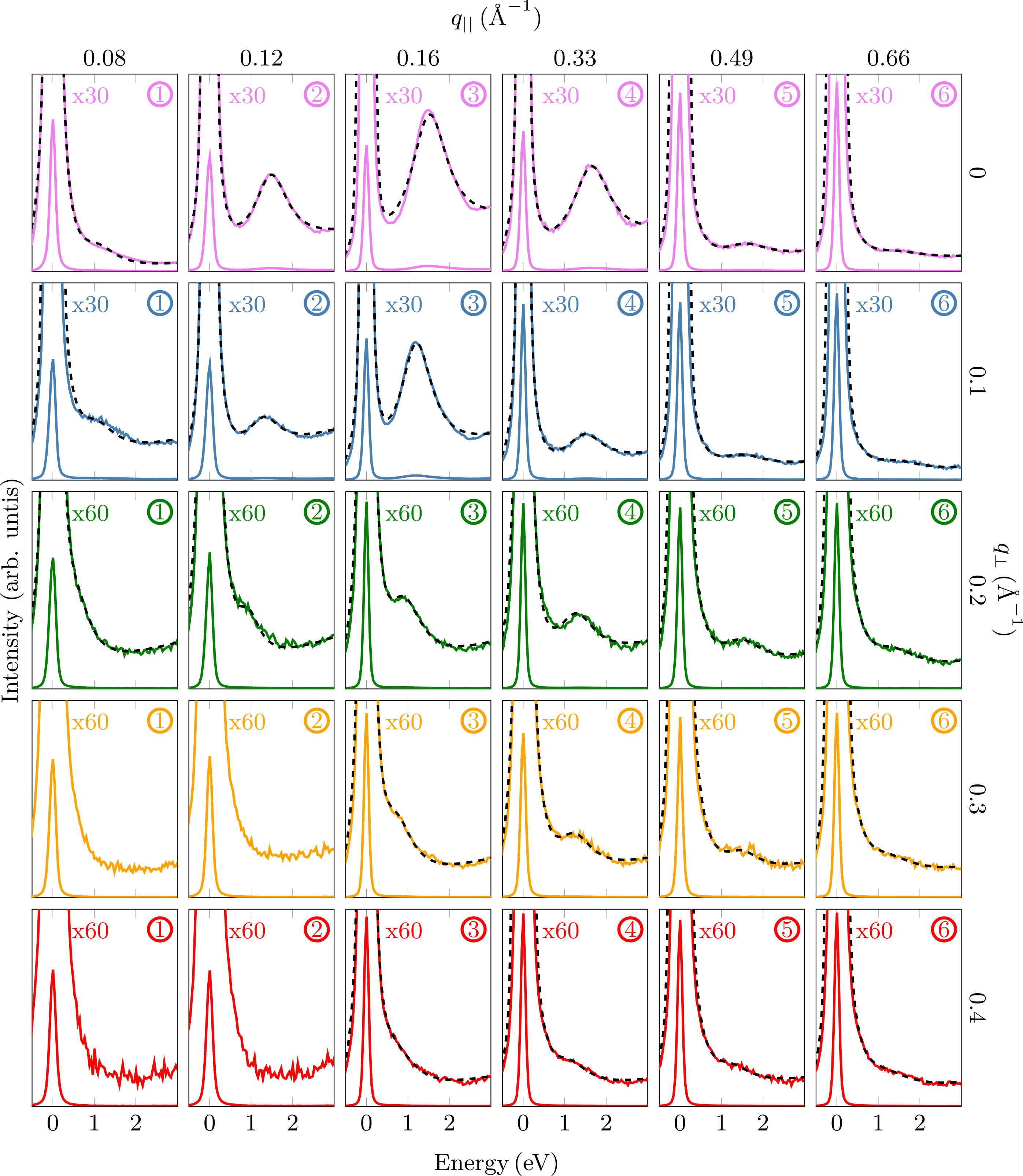}
		\caption{Electron energy-loss intensities for various $q_{||}$ and $q_{\perp}$ values. The indicated numbers (1-6) and the colors correspond to the momentum values depicted in Fig.\,\ref{fig:fig2}. Experimental spectra are depicted twice (normal scale + 30x/60x scaled up) to show both the zero loss peak and the plasmon peak. The black dashed lines correspond to fits of a superposition of a Voigt profile (zero loss), a Drude function (plasmons), and a background (see Methods).} 
		
		\label{fig:fig3}
	\end{figure*}

	The data presented in Fig.~\ref{fig:fig3} demonstrate that T-EELS in layered systems is capable of probing not only  in-phase (optical) plasmons with momentum parallel to the layers but also   out-of phase (acoustic) collective charge oscillations in neighboring layers. A suitable sample preparation is important for such studies.
	Thus, we show that T-EELS can compete with recent RIXS studies of acoustic plasmons of cuprates~\cite{Hepting2018,Nag2020,Hepting2022}. This is an important extension of T-EEL spectroscopy.
	Furthermore, we emphasize that the present work shows that acoustic plasmons exist also in non-cuprate "strange" metal layer systems. 
	\par
	Well defined plasmons exist in the complete range which is not covered by the range of single-particle particle excitations calculated in the mean-field RPA theory. Optical plasmons exist in $ \approx $ 15 \% of the BZ . The rest is determined by a continuum of intra-band single particle excitations, which strongly dampen the plasmon excitation in excellent agreement with previous studies~\cite{Knupfer2022}. There is no sign of a reduction of the coherent plasmon range due to an over-damping discussed in terms of holographic theories~\cite{Romero-Bermudez2019,eede2023}.
	\par
\begin{figure}[ht!]
		\centering
		\includegraphics[width=0.475\textwidth]{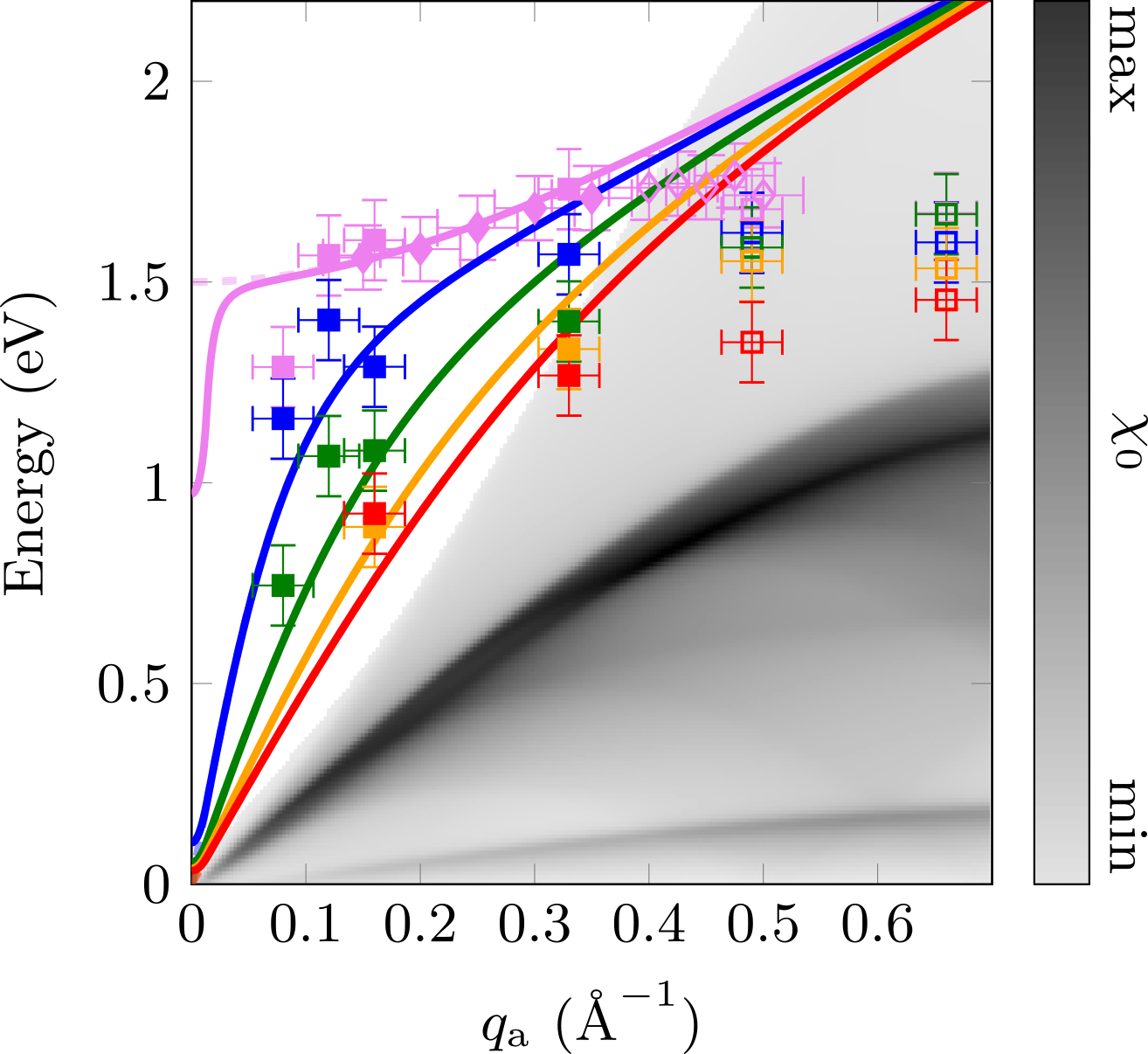}
		\caption{Plasmon dispersion along the momentum $q_{||}$ parallel to the layers for several $q_\perp$ perpendicular to the layers (squares) together with calculations within the framework of the Fetter model (solid lines). The $q_\perp$ values 0, 0.1, 0.2, 0.3, and 0.4 \AA $^{-1}$ are marked by purple, blue, green, yellow, and red colour {symbols}, respectively (see Fig.~\ref{fig:fig3}). We have added also the data from the optical plasmon dispersion derived in our previous publication dark purple stars)~\cite{Knupfer2022}. The horizontal error bars originate from the finite momentum resolution while the vertical ones are related to the finite spectral resolution of the EEL spectrometer ( the fitting error is small in comparison).
			The region marked in grey corresponds to the susceptibility $\chi_0$ calculated from a tight binding band structure (see below and Methods). The excitations in the continuum range are marked by open symbols. The colored dashed curves represent theoretical curves considering finite momentum resolution, i.e., integration within the EEL collection aperture.}
		
		\label{fig:fig4}
	\end{figure}

	Fitting the dispersion of the optical plasmon dispersion for $q_{||}\le q_\mathrm{crit} \approx $ 0.4 \AA $^{-1}$ (see Fig.~\ref{fig:fig4} purple data) with Eq.~(\ref{eq:dis}) we derive a dispersion coefficient $A=$ 1.8 $\pm$ 0.8 eV\AA$^2$ which is in agreement with the more precise result from our previous publication~\cite{Knupfer2022} $A=$ 2.1 $\pm$ 0.2 eV\AA$^2$. In the following we discuss the latter $A$ value from our previous publication. 
	\par

	\begin{figure*}[ht!]
		\centering
		\includegraphics[width=.95\textwidth]{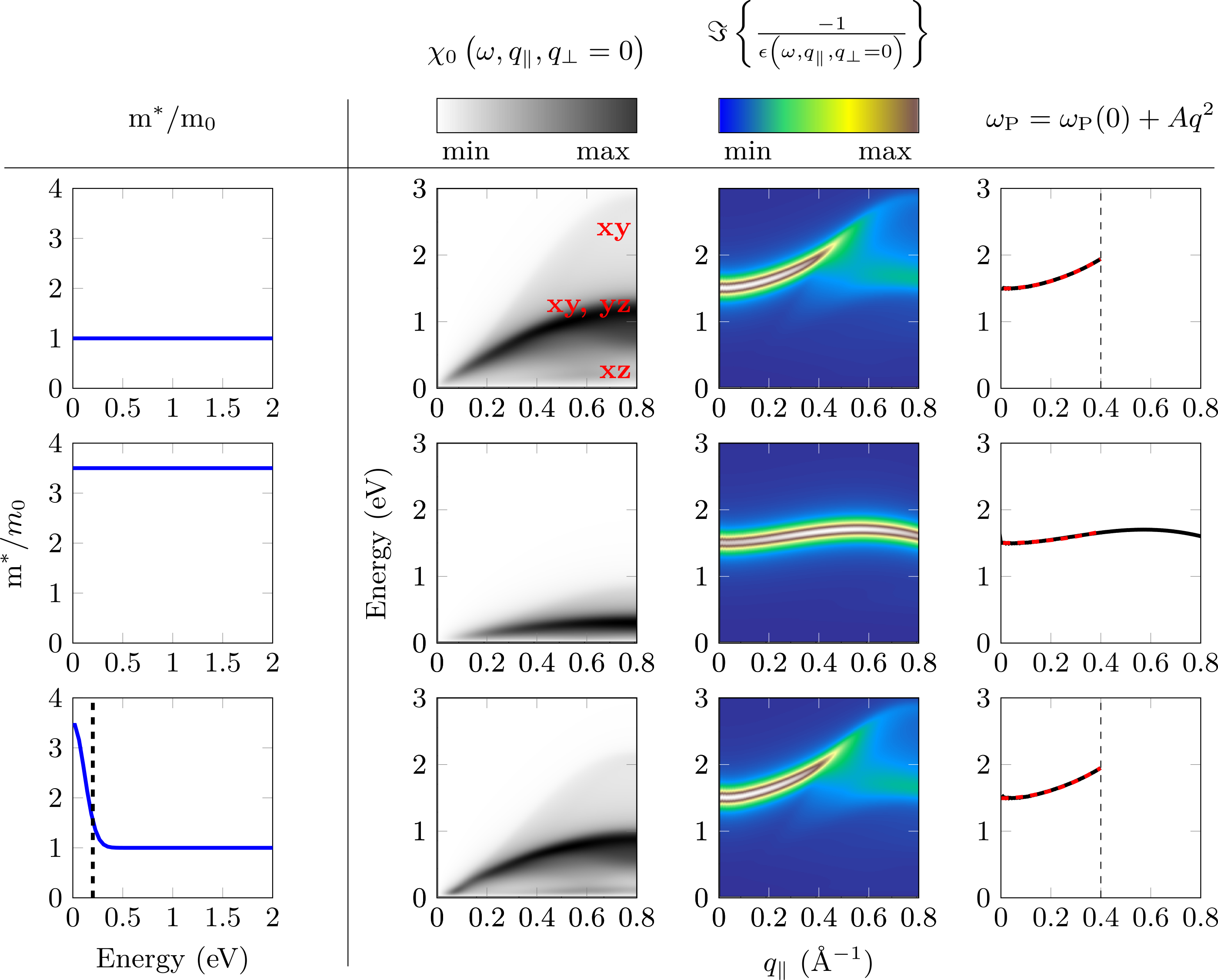}
		\caption{Calculated imaginary part of the susceptibility $\chi_0 (\omega, q_{||},q_{\perp}=0)$ (second column), loss function $\Im \left\lbrace-\frac {1}{\epsilon(\omega, q_{||}, q_{\perp}=0)}\right\rbrace$ (third column), and plasmon dispersion (fourth column) along the momentum $(q_{||},q_{\perp}=0)$ together with least squares fit (red lines) in the momentum range $q_{||}$= 0 to 0.4 \AA $^{-1}$. In the calculation various effective masses were used: first row $m^*/m_0=$1, second row $m^*/m_0=$3.5, and an energy dependent effective mass $m^{\ast}(\omega)/m_0$ shown in the third row. The red labels in the susceptibility in the first row are related to the origin of the three electronic bands in Sr$_2$RuO$_4$. }
		\label{fig:calc}
	\end{figure*}
	\begin{figure}[t]
		\centering
		\includegraphics[width=0.475\textwidth]{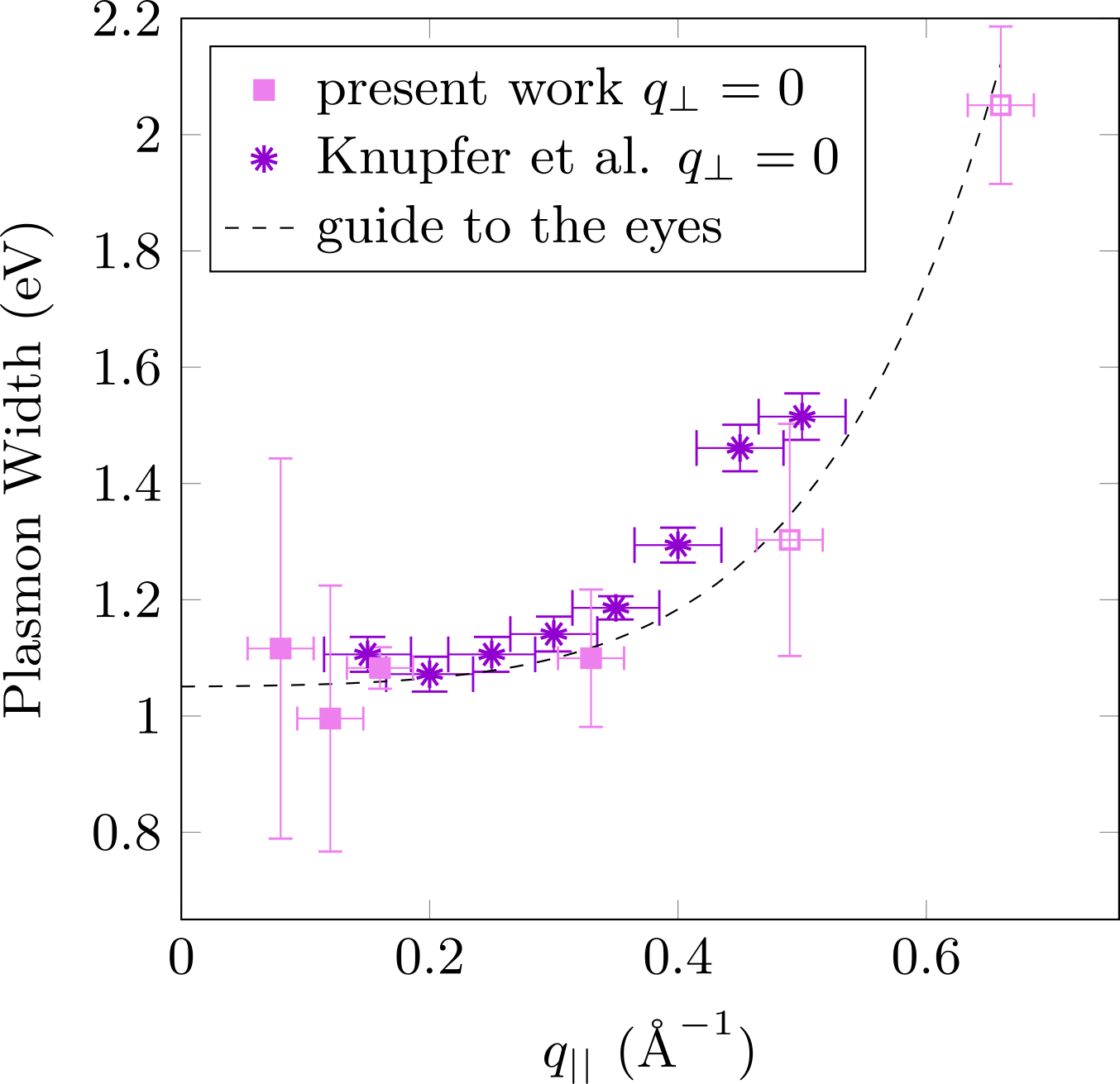}
		\caption{Plasmon 
			width at half maximum in dependence on the momentum transfer parallel to the layers at $q_\perp=0$. For comparison the corresponding data from a Drude fit of our previous EELS measurements reported in Ref. ~\cite{Knupfer2022} was added as dark purple stars. The energy of the maximum of the excitations in the continuum range (above $q_{||} $ are marked by open symbols.} 
		\label{fig:width}
	\end{figure}
	Next we compare the present EELS data of the optical plasmon dispersion with those derived from theoretical calculations of the loss function. 
		In Fig.\,\ref{fig:calc} we present these results for the susceptibility, the loss function, and the optical plasmon dispersion for an unrenormalised band structure~\cite{Liebsch2000} ($m^*/m_0=$1), a constant effective mass ($m^*/m_0=$ 3.5), and an energy dependent effective mass ranging from 3.5 at low energies to 1 above $\approx $ 0.2 eV. The latter was taken from optical spectroscopy~\cite{Stricker2014} together with an extrapolation to 1 at higher energies.
	\par
	Using the unrenormalised tight binding band structure~\cite{Liebsch2000} we receive below $q_{\text{crit}}$ the dispersion coefficient $A = $ 2.8 eV \AA $^2$. The latter value is close to the experimental value $A= 2.1 \pm 0.2$ eV\AA$^2$. The small difference between the experimental and the theoretical $A$ value can be explained by a finite thickness of the RuO$_2$ layers which reduces the nominal $d$ to $d_\mathrm{reduced}=$ 4. This reduction  of $d$ brings the theoretical $A$ value very close to the experimental one. Moreover, our calculations show that an enlargement of the  half width with $\Gamma$ from about 0.1 to  the experiment reduces the dispersion coefficient $A$ by $\approx 0.2 $\,eV\AA$^2$.
	
	\par 
	
		On the other hand, using the constant effective mass $m^*/m_0=$ 3.5 from low-energy/low temperature optical spectroscopy we derive a reduced dispersion with $A=$ 1.1 eV\AA$^2$ which is at variance with the experimental data. This effective mass dependence of the plasmon dispersion was already discussed in our previous paper~\cite{Knupfer2022}. We also remark that the susceptibility is strongly reduced in energy compared to the plasmon energy. Therefore in case of $m^*/m_0=$ 3.5 the plasmon is not merging into a continuum and exists well above $q_\mathrm{crit}$. 
	
	\par
	Using the energy dependent effective mass from optical spectroscopy below $\omega=$ 0.2 eV plus an extrapolation to $m^*(\omega)/m_0=1$ at higher energies we see the expected strong renormalization and an intensity increase in the low energy/momentum range. However, at higher energy/momentum the  susceptibility is similar to that calculated with an unrenormalised band structure. This causes a plasmon dispersion which is very close to an unrenormalised dispersion. Thus our present and the previous experimental data of the dispersion of the optical plasmon indicates that the long wavelength dispersion can be explained in the framework of a mean-field RPA theory using an effective mass of one. Because for the cuprates an energy dependent effective mass is expected as well, probably the present result can be also explain the unrenormalised plasmon dispersion detected in the cuprates~\cite{Nuecker1991,Grigoryan1999}.
	
	\par
	The calculations  clearly demonstrate that the low energy renormalization of the susceptibility/optically conductivity does not transfer into the high-energy plasmon dispersion. On the other hand, the renormalised acoustic plasmon dispersion observed by RIXS at low energies in cuprates~\cite{Nag2024} can be explained in this framework.

	\par
	In Fig.~\ref{fig:width} we present the optical plasmon width as a function of $q_{||}$. We show  data derived from 
	Fig.~\ref{fig:fig3} and from our previous EELS 
	experiments~\cite{Knupfer2022}. Within error bars there is a good agreement between the two datasets. The plasmon width at zero momentum is smaller than the plasmon energy, indicating a coherent collective charge excitation. The width below  $q_{||} \approx 0.4$ \AA $^{-1}$  is constant. 
	Near $q_{||} = $0.4 \AA $^{-1}$ the width increases, indicating the merging of the plasmon dispersion into the single particle continuum at a $q_\mathrm{crit}\approx $ 0.4 \AA $^{-1}$. (See also Fig. \ref{fig:fig4}). As the calculations for the continuum are performed without band-renormalisation  this indicates again that at the relative high plasmon energy, the effective mass is close to one and supports the formation of resilient quasi-particles~\cite{Stricker2014}.
	Already in our previous paper~\cite{Knupfer2022}, we have emphasised that the width is almost constant in the studied low momentum range. This means that it is not related to electron-electron interaction which would lead to a quadratic increase of the width as a function of momentum transfer~\cite{DuBois1969}. The fact that the width is smaller than the energy signals that the excitation is not overdamped due to fluctuations in a quantum critical system, in contrast to theoretical predictions~\cite{Romero-Bermudez2019}.
		Rather, it is caused, as in most metallic systems, studied by T-EELS, by a decay into interband transitions~\cite{Paasch1970,Gibbons1977,Felde1989a,Sturm2000}. These interbands are induced by a back-folding of bands from the second to the first Brillouin zone by a finite pseudo-potential. In this way, in recent RIXS data on p and n-typed cuprates~\cite{Nag2024}, the difference in accoustic plasmon width could be described in this framework. 
	
	\par
\begin{figure*}[!ht]
		\centering
		\includegraphics[width=0.95\textwidth]{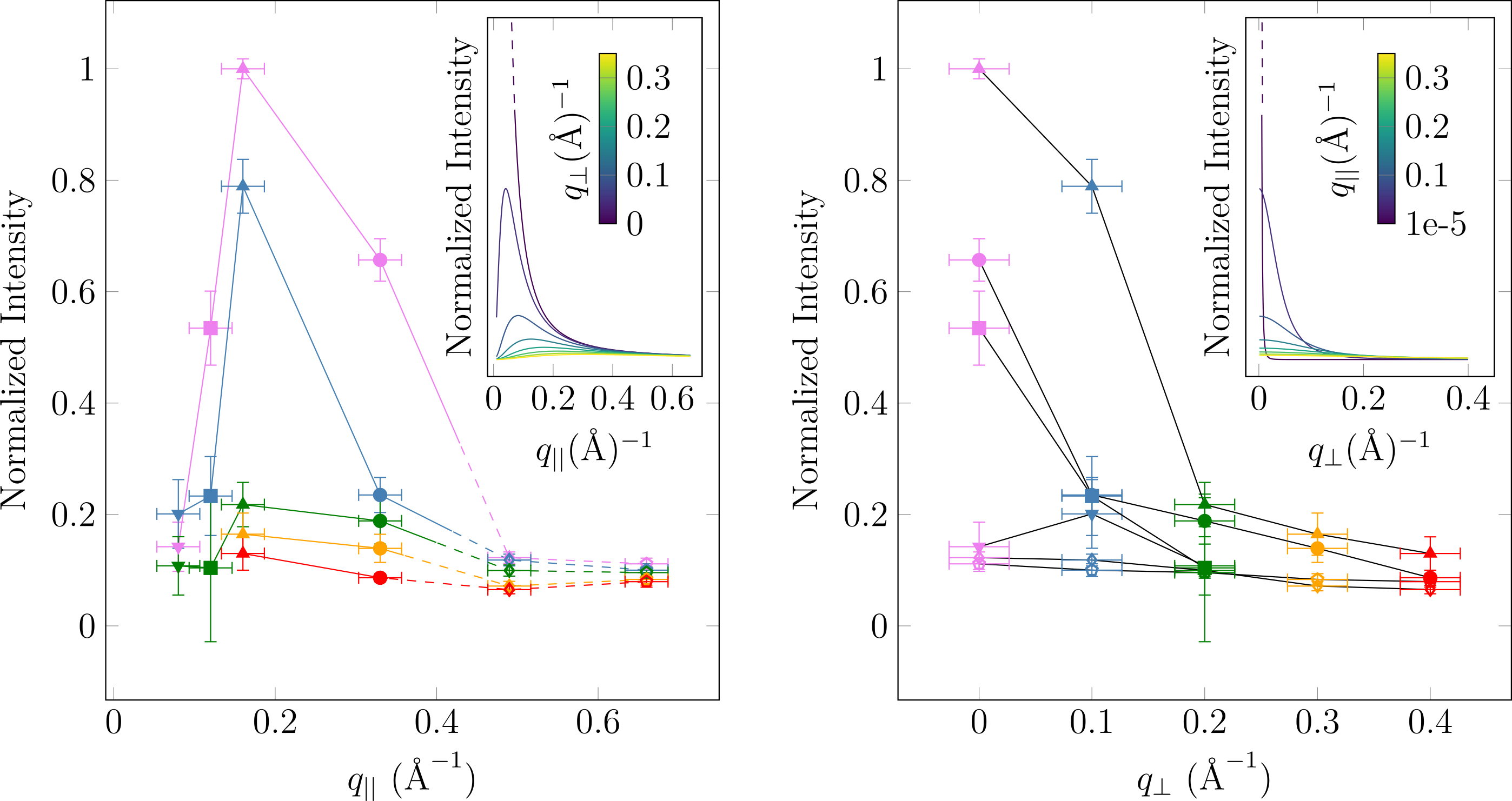}
		\caption{Integrated loss intensities under the fitted plasmon peaks as a function of in-plane ($q_{||}$) and out-of-plane ($q_\perp$) momentum transfer. The $q_\perp$ values 0, 0.1, 0.2, 0.3, and 0.4 \AA$^{-1}$ are marked by purple, blue, green, yellow, and red colour symbols, respectively (see Fig. 3). The $q_{||}$ values 0.08, 0.12, 0.16, 0.33, 0.49, and 0.66 \AA$^{-1}$ are marked by tip down triangles, squares, tip top triangles, circles, diamonds, and pentagons, respectively. The dashed lines and open symbols indicate the single particle continuum regime. The insets show the dependency predicted by the coupled 2D plasmon model due to Ref.\,\cite{Fetter1974}. At $q_{\perp}=q_{||}=0$ the curves diverge indicated by a dashed line.} 
		\label{fig:intensities}
	\end{figure*}

	In Fig.~\ref{fig:intensities} we depict the plasmon intensities as a function of $q_{||}$ and  $q_{\perp}$ (see Methods section for details of the evaluation).
	The decay of the total spectral weight of the plasmon resonances at large momentum transfers  is approximately proportional to $q^{-2}$ as observed for conventional bulk plasmons and predicted by the longitudinal f-sum rule and the theoretical dielectric function of the Fetter model (see Methods). The decay observed for $q_\perp\ne0$ in the long wavelength limit is also consistent with the theoretical dielectric response as well as a version of the long wavelength sum rule implying that the full spectral weight at $q=0$ is concentrated in the bulk plasmon $q_{\perp}=0$ mode (see Methods and Ref. \cite{mahan2000} for the sum rules). The shift of the maximal integrated loss intensity towards smaller momentum transfer predicted by theory may be attributed to shortcomings of the Fetter model dielectric function (see Methods), impact of the zero loss, and limited experimental resolution.
	\par
	In the following we discuss the acoustic plasmon data (see red data and line in Fig.~\ref{fig:fig4}). Despite the gap at low energy due to a finite energy resolution, the dispersion extrapolates to zero energy typical of an acoustic plasmon. The derived experimental plasmon velocity is $v_\mathrm{P} \approx$ 4.7 eV\AA.
	From Eq.~(\ref{eq:acoust}) we derive $v_\mathrm{P} \approx $ 4.8 eV\AA \hspace{0.2cm} in very good agreement with the experimental value. This indicates that the first term in Eq.~(\ref{eq:acoust}) due to the finite Fermi velocity is small compared to the term which only depends on $\omega_\mathrm{P}(0)$ and $d$. Thus, for a given 
	$\omega_\mathrm{P}(0)$ the acoustic plasmon dispersion only depends on $d$ and is not influenced by a large Fermi velocity but hints to a reduced one due to correlation effects (enhanced effective mass). Unfortunately, the finite energy resolution in the present T-EELS experiment does not allow a quantitative 
	determination of the effective mass at low energies. However, low-energy RIXS data on cuprates indicate, that an enhanced effective mass is necessary to describe the plasmon velocity of the acoustic plasmons~\cite{Nag2024}. 
	\par
	Neglecting the Fermi velocity term, the linear acoustic plasmon dispersion [see Eq.~(\ref{eq:acoust})] can be explained in the following way. The phase difference $\pi$ of the oscillations between neighboring layers reduces the plasmon energy from $\omega_\mathrm{P}(0)$ to zero. When adding a momentum $q_{||}$ the phase difference between neighboring layers is increased to $\pi + q_{||} d$ and therefore, using a linear relation, the energy of the acoustic plasmon should increase by $(\omega_\mathrm{p}(0) d q_{||})/\pi$
	which is close to Eq.~(\ref{eq:velocity}).
	\par
	In summary, the long wavelength $q_{||}$ dispersion of the $q_\perp$ dependent plasmons including the optical and the acoustic collective excitations and the decay of the optical plasmon by Landau damping
	can be all explained in terms of a mean-field RPA model. It is possible to understand this interpretation of the present results in the following way. Long wavelength charge excitations are not influenced by local interactions such as on-site Coulomb and Hund's exchange interactions. This behaviour is different from ARPES studies, in which local properties play an important role. In this context it is also important to note that monopole (a single hole) excitations detected in ARPES are differently screened compared to dipole excitations recorded in EELS. We further emphasize that our present analysis of the acoustic plasmon dispersion is also important for the understanding of previous~\cite{Hepting2018,Nag2020,Hepting2022} and future RIXS studies on cuprates.
	
	\section{Perspectives}
	The present study has demonstrated that optical and acoustic plasmons can be investigated by T-EELS in the complete BZ in layered systems. Therefore, with the advent of higher energy resolution, T-EELS will be competitive at lower energies with RIXS, also taking into account that momentum-resolved T-EELS provides a direct probe of the dynamic susceptibility. 
	It will be possible to study in more detail the different influence of correlation effects on optical and acoustic plasmons, caused by an energy dependent effective mass. The latter was predicted by a combined density functional/ dynamical mean-field theory (DFT + DMFT) calculation~\cite{Deng2013} and experimentally detected by optical spectroscopy~\cite{Stricker2014}. Furthermore, in "strange" metals, it will be possible to study low-energy and high-momentum charge excitations which were predicted in Ref.~\cite{Khaliullin1996} to depend on correlation effects. In this way it will be possible to evaluate the spatial dependence of the density-density fluctuations in "strange" metals.
	\section{Methods}

	\subsection{Dielectric Response of the Layered Plasmon System}
	
	The dielectric function corresponding to the Fetter model of a system of coupled 2D layers reads \cite{Fetter1974}\\
	\vspace{-0.1cm}
	\begin{eqnarray}
		&&\epsilon(\omega,q_{||},q_{\perp})=\nonumber\\
		&&1-\frac{2\pi Ne^2q_{||}/m}{\omega\left(\omega+i\Gamma\right)-s^2q_{||}^2}\frac{\sinh(q_{||}d)}{\cosh(q_{||}d)-\cos(q_{\perp}d)}.\nonumber\\
	\end{eqnarray}
	This dielectric function is an approximation assuming a perturbation charge that is confined to the 2D layers supporting the plasmons, which is violated by the electron beam resulting in deviations to the experimental dielectric response. However, the intensities of the plasmon peaks can be derived from this dielectric function by calculating the loss probability (T-EELS signal) using Eq. \ref{eq:dis} and integration along $\omega$ (see Fig. \ref{fig:intensities}).\\
	A version of the long wave wavelength sum rule for the dynamic susceptibility reads \cite{mahan2000}
	\begin{equation}
		\underset{q\rightarrow0}{\lim}\Im\left(\frac{1}{\epsilon\left(q,\omega\right)}\right)=-\frac{\pi\omega_{\text{P}}}{2}\left(\delta\left(\omega-\omega_{\text{P}}\right)-\delta\left(\omega+\omega_{\text{P}}\right)\right).
	\end{equation}
	Thus, in the $q\to 0$ limit, the dynamic susceptibility is determined by the longitudinal plasmon.
	\subsection{Samples}
	Sr$_2$RuO$_4$ crystals were grown using the  floating-zone
	method~\cite{Bobowski2019}. The superconducting transition temperature of the sample was $T_\mathrm{c}=$1.5 K.
	\par
	Thin TEM lamellas of Sr$_2$RuO$_4$ with the normal pointing along 
	${\bf b}$ 
	were prepared by Focused Ion Beam (FIB) using a Thermofisher  
	instrument. The target thickness of the lamellas was 80 nanometers. Low ion energy polishing was used as final step to thin Ga-ion damaged surface layers.
	\subsection{EELS measurements}
	The momentum-resolved loss function 
	was recorded at a FEI Titan$^3$ TEM equipped with a Wien monochromator and a Gatan Tridiem imaging filter (GIF) at 80 kV acceleration voltage in a serial way (i.e., one EEL spectrum per fixed momentum). The energy resolution is around 120\,meV (FWHM of zero loss peak). At a camera length (i.e., effective distance between sample and detector) of 1.15\,m, the GIF entrance aperture was used to select the different momentum values and covers a momentum range of 0.04\,$\Aa^{-1}$ (0.13\, mrad semi collection angle). The acquisition times are presented in Table 1.
	\begin{table}[!ht]
		\caption{
			Acquisition times (in sec) for the used momentum transfers (in \AA  $^{-1}$).}
		\begin{tabular}{c|c|c|c|c|c|cc} 
			\backslashbox{$q_{\perp}$}{$q_{||}$}
			0.08
			& 0.12
			& 0.16
			&0.33
			& 0.49
			& 0.66
			&\\
			\hline
			0
			& 2
			& 5 
			& 5 
			& 5 
			& 10
			& 10 
			&\\
			0.1
			& 5  
			& 5 
			& 5 
			& 5 
			& 10 
			& 10
			& \\
			0.1
			& 5  
			& 5
			& 5 
			& 5 
			& 10
			& 10
			& \\
			0.3
			& 10
			& 10
			& 10
			& 10
			& 10
			& 10
			& \\ 
			0.4
			& 10  
			& 10 
			& 10
			& 10 
			& 10 
			& 10
			& .\\
		\end{tabular}
	\end{table}
	\\
	Due to instabilities of the monochromator and the rather long collection times required at large momentum transfers, the recorded spectra are subject to substantial noise as well as mutual random fluctuations/offsets. In order to mitigate these effects, each spectrum was separately aligned with the help of the quasi-elastic peak. After alignment of the zero loss position a superposition of the following three functions was fitted to the spectra. i) an asymmetric Pseudo-Voigt-profile ($V=c\left(\eta \frac{1}{1+(\omega-\omega_0)^2/(\lambda(\omega) \sigma)^2}+\left(1-\eta\right)\exp\left[-\ln(2)\left(\frac{\omega-\omega_{0}}{\sigma}\right)^2\right]\right)$ with $\lambda(\omega)=1|_{\omega<=0}$ and $\lambda(\omega)>1|_{\omega>0}$) reflecting the quasi elastic peak including ultra-low-loss excitations such as phonons, ii) a Drude-like function $I\left(\omega\right)=a\frac{\omega_{\text{P}}^2\omega\Gamma}{\left(\omega^2-\omega_{\text{P}}^2\right)^2+\left(\omega\Gamma\right)^2}$ corresponding to the plasmon peaks and iii) a phenomenological background following a linear function. Finally, the spectral positions and half widths of the plasmons were derived from the fitted parameters $\omega_{\text{P}}$ and $\Gamma$ of the Drude function. The intensity (Fig. \ref{fig:intensities}) corresponds to the area under the fitted Drude peak, which is obtained by integrating the latter over $\omega$.
	
	\subsection{Calculation of the average Fermi velocity, susceptibility, and loss function}

		For the calculation of the Fermi velocity along the [100] direction, the bare particle susceptibility $\chi_0 (q_{||},\omega)$ and the loss function we used a TB band structure~\cite{Liebsch2000} based on an LDA band calculation. $\chi_0 (q_{||},\omega)$ is calculated from a  multi-band version of Eq.~\ref{eq:Ehren}, taking only intra-band excitations with the same matrix element into account, thus neglecting inter-band excitations which may be caused by a strong spin-orbit coupling, which leads to a $k$-dependence of the orbital character of the bands~\cite{Tamai2019}.
		Mainly the XY and the YZ bands contribute to $\left<v_\text{F}^2\right>_{100}$ and $\chi_0$.
		The dielectric function and the loss function were calculated from Eqs.~\ref{eq:dis}, \ref{eq:Ehren}, and \ref{eq:diel}.  $\epsilon_{\text{b}}$ and the matrix element were fixed by using the plasmon energy from optical spectroscopy.  For the calculation we used a half width of $\Gamma  =$ 0.1 eV.

	\section{Data availability}
	All data supporting the findings are provided as figures in the article. Data files for all figures are available at https://opara.zih.tu-dresden.de//handle/123456789/763 and from the corresponding authors on request.
	
	\section{References}
	\bibliography{Sr2RuO4.bib}

	\begin{acknowledgments}
		J.F. thanks P. Abbamonte, R. von Baltz, S.-L. Drechsler, and A. Greco for helpful discussions. This work is supported by a KAKENHI Grants-in-Aids for Scientific Research (Grants No. 18K04715, No. 21H01033, and No. 22K19093), and Core-to-Core Program (No. JPJSCCA20170002) from the Japan Society for the Promotion of Science (JSPS) and by a JST-Mirai Program (Grant No. JPMJMI18A3). J.S. received funding from the HORIZON EUROPE framework program for research and innovation under grant agreement n. 101094299. A.L. and M.K. acknowledge funding from the Deutsche Forschungsgemeinschaft (DFG, German Research Foundation)-project-id 461150024. B.B. received funding from the Würzburg-Dresden Cluster of Excellence on Complexity and Topology in Quantum Matter – ct.qmat (EXC 2147, project-id 390858490).
		
	\end{acknowledgments}
	
	\section{Author contributions}
	J.F., M.K., A.L., and B.B. conceived the experiment, J.S. and D.W. performed the EELS experiment. J.S. and J.F. analysed the data.
	F.J. and N.K. prepared and characterised the samples. J.F. calculated the susceptibility and the loss function. J.F., J.S, and A.L. wrote the manuscript with input from all authors.
\end{document}